\documentclass[10pt,conference]{IEEEtran}
\IEEEoverridecommandlockouts
\usepackage{cite}
\usepackage{amsmath,amssymb,amsfonts}
\usepackage{algorithmic}
\usepackage{graphicx}
\usepackage{textcomp}
\usepackage{xcolor}
\def\BibTeX{{\rm B\kern-.05em{\sc i\kern-.025em b}\kern-.08em
    T\kern-.1667em\lower.7ex\hbox{E}\kern-.125emX}}

\usepackage{tikz}
\usepackage{graphicx}
\usetikzlibrary{arrows.meta,positioning,fit,backgrounds,calc}
\usepackage{todonotes}
\usepackage{cleveref}
\usepackage{listings}
\usepackage{caption}
\usepackage{url}
\usepackage{booktabs}

\DeclareCaptionFormat{listing}{\vskip 6pt #1#2#3}

\usepackage{xspace}

\newcommand{\tool}{\texttt{MLIR-MRacle}\xspace{}}

\newcommand{\mlir}{\texttt{MLIR}\xspace{}}
\newcommand{\llvm}{\texttt{LLVM}\xspace{}}
\newcommand{\openmp}{\texttt{OpenMP}\xspace{}}
\newcommand{\async}{\texttt{async}\xspace{}}
\newcommand{\libafl}{\texttt{libafl}\xspace{}}
\newcommand{\cpp}{\texttt{C++}\xspace{}}
\newcommand{\clang}{\texttt{C/C11}\xspace{}}
\newcommand{\rust}{\texttt{Rust}\xspace{}}

\newcommand{\arm}{\texttt{Armv8}\xspace}

\begin{document}

\title{\tool{}: Memory-Model-Aware Metamorphic Fuzzing of Concurrent \mlir{}\\
\thanks{Aidan Dakhama is supported by the UKRI Centre for Doctoral Training in Machine Learning Systems, funded by EPSRC grant EP/Y03516X/1.}
}

\author{
  \IEEEauthorblockN{Aaron Gabayoyo}
  \IEEEauthorblockA{\textit{School of Informatics} \\
    \textit{University of Edinburgh}\\
    Edinburgh, United Kingdom \\
    A.Gabayoyo@sms.ed.ac.uk
  }
  \and
  \IEEEauthorblockN{Aidan Dakhama}
  \IEEEauthorblockA{\textit{School of Informatics} \\
    \textit{University of Edinburgh}\\
    Edinburgh, United Kingdom \\
    aidan.dakhama@ed.ac.uk
  }
  \and
  \IEEEauthorblockN{Ajitha Rajan}
  \IEEEauthorblockA{\textit{School of Informatics} \\
    \textit{University of Edinburgh}\\
    Edinburgh, United Kingdom \\
    arajan@ed.ac.uk
  }
}

\maketitle

\begin{abstract}
Multi-level compilers built on \mlir{} lower concurrent programs through many stages towards targets with potentially drastically different memory models, where an optimisation which appears reasonable, can silently change program behaviour. 
Testing these lowerings is hard: a concurrent program often has no single expected outcome, and existing \mlir{} fuzzers often assume determinism. 
We present \tool{}, a memory-model-aware metamorphic fuzzer for concurrent \mlir{} programs. 
\tool{} applies metamorphic relations gated by the target memory model, so that both the original and transformed programs remain well defined -- executing both repeatedly, and comparing their outcome sets or distributions with one of two statistical oracles.
A coverage guided campaign driver built on \libafl{} is used to perform continuous fuzzing. 
MRs, memory models, oracles, and execution backends all provide extension points for future researchers and compiler engineers to build off this package.
In a case study on \arm{} hardware, \tool{} ran over 145k trials across four configurations, reaching up to 93.6k compiler coverage edges, and found a bug in the \openmp{} dialect verifier that has since been fixed upstream.
Video: \url{https://youtu.be/lKtGk0H97Ts?si=dWLUHl2CJ24ep6JY}
\end{abstract}


\section{Introduction}

\definecolor{drvfill}{HTML}{DCE9F5}\definecolor{drvline}{HTML}{2B5C8A}
\definecolor{engfill}{HTML}{F6E7D3}\definecolor{engline}{HTML}{9A5B13}
\definecolor{extfill}{HTML}{D6EFE8}\definecolor{extline}{HTML}{1F7A63}

\begin{figure}[hbt!]
\centering
\resizebox{\columnwidth}{!}{%
\begin{tikzpicture}[
  font=\footnotesize,
  box/.style={draw, rounded corners=1.5pt, align=center, line width=0.5pt,
              minimum width=29mm, minimum height=7.5mm, inner sep=2pt},
  drv/.style={box, fill=drvfill, draw=drvline},
  eng/.style={box, fill=engfill, draw=engline},
  ext/.style={draw=extline, fill=extfill, rounded corners=1.2pt, align=center,
              line width=0.45pt, minimum width=14mm, minimum height=6.5mm,
              inner sep=1pt, font=\scriptsize},
  fut/.style={ext, dashed, fill=white},
  grp/.style={draw=extline!60, rounded corners=2pt, line width=0.4pt,
              inner xsep=2.5pt, inner ysep=2pt},
  ghdr/.style={font=\scriptsize\itshape, text=extline, anchor=south west,
               inner sep=1pt},
  sub/.style={font=\scriptsize},
  arr/.style={-Latex, line width=0.5pt},
  plug/.style={-Latex, line width=0.5pt, draw=extline},
]

\node[drv] (corpus) at (0, 0)     {Seed corpus};
\node[drv] (sched)  at (0,-1.15)  {Trial scheduling\\[-1pt]{\scriptsize seed, MR, model,}\\[-2pt]{\scriptsize backend, oracle}};
\node[eng] (mr)     at (0,-2.8)   {MR application\\[-1pt]{\scriptsize memory-model gated}};
\node[eng] (agit)   at (0,-3.92)  {Lowering\\[-1pt]{\scriptsize + agitation (outcome only)}};
\node[eng] (exec)   at (0,-5.04)  {Execution};
\node[eng] (oracle) at (0,-6.16)  {Oracle verdict};
\node[drv] (triage) at (0,-7.76)  {Feedback, triage\\[-1pt]{\scriptsize archival \& replay}};

\foreach \a/\b in {corpus/sched, sched/mr, mr/agit, agit/exec, exec/oracle, oracle/triage}
  \draw[arr] (\a) -- (\b);

\node[ext] (gen)  at (2.6,-2.8)   {generic};
\node[ext] (arm)  at (4.05,-2.8)  {Armv8};
\node[fut] (omm)  at (5.5,-2.8)   {other\\[-2pt]models};
\node[grp, fit=(gen)(omm)] (gmr) {};
\node[ghdr] at (gmr.north west) {MR plug-ins};

\node[ext] (jit)  at (2.6,-5.04)  {LLJIT};
\node[ext] (nat)  at (4.05,-5.04) {native\\[-2pt]process};
\node[fut] (sim)  at (5.5,-5.04)  {hardware /\\[-2pt]simulator};
\node[grp, fit=(jit)(sim)] (gbe) {};
\node[ghdr] at (gbe.north west) {backend adapters};
\node[ext] (ago) at (2.6,-6.16)  {outcome};
\node[ext] (cal) at (4.05,-6.16) {adaptive};
\node[grp, fit=(ago)(cal)] (gor) {};
\node[ghdr] at (gor.north west) {oracle modes};

\draw[plug] (gmr.west) -- (mr.east);
\draw[plug] (gbe.west) -- (exec.east);
\draw[plug] (gor.west) -- (oracle.east);

\begin{scope}[on background layer]
  \coordinate (etop) at ($(gmr.north)+(0,0.38)$);
  \coordinate (ebot) at ($(oracle.south)+(0,-0.45)$);
  \node[fit=(mr)(gmr)(gor)(etop)(ebot), inner xsep=4pt, inner ysep=0pt,
        fill=engfill!35, draw=engline!50, rounded corners=3pt,
        dash pattern=on 2pt off 1pt] (engine) {};
\end{scope}
\node[sub, text=engline, anchor=south east, inner sep=2pt]
  at (engine.south east) {\textbf{native engine (C++)}};
\node[sub, text=drvline, anchor=west, font=\scriptsize\bfseries]
  at ($(corpus.east)+(0.2,0)$) {campaign driver (Rust, \libafl{})};

\draw[arr, draw=drvline]
  (triage.west) -- ++(-0.4,0) |- (corpus.west)
  node[pos=0.25, sloped, above, sub, text=drvline, inner sep=1.5pt]
  {coverage + semantic feedback};

\node[ext, minimum width=4.5mm, minimum height=3mm] (l1) at (1.95,-7.76) {};
\node[sub, anchor=west] at (l1.east) {implemented};
\node[fut, minimum width=4.5mm, minimum height=3mm] (l2) at (4.4,-7.76) {};
\node[sub, anchor=west] at (l2.east) {extension point};
\end{tikzpicture}%
}
\caption{\tool{} architecture. A campaign driver schedules trials over a corpus; each trial calls the native engine. Teal boxes are pluggable: MRs, execution backends and oracle modes are selected independently per trial; agitation applies only under the outcome oracle. Dashed boxes are extension points, e.g.\ memory models or a simulator backend.}
\label{fig:mracle-architecture}
\end{figure}
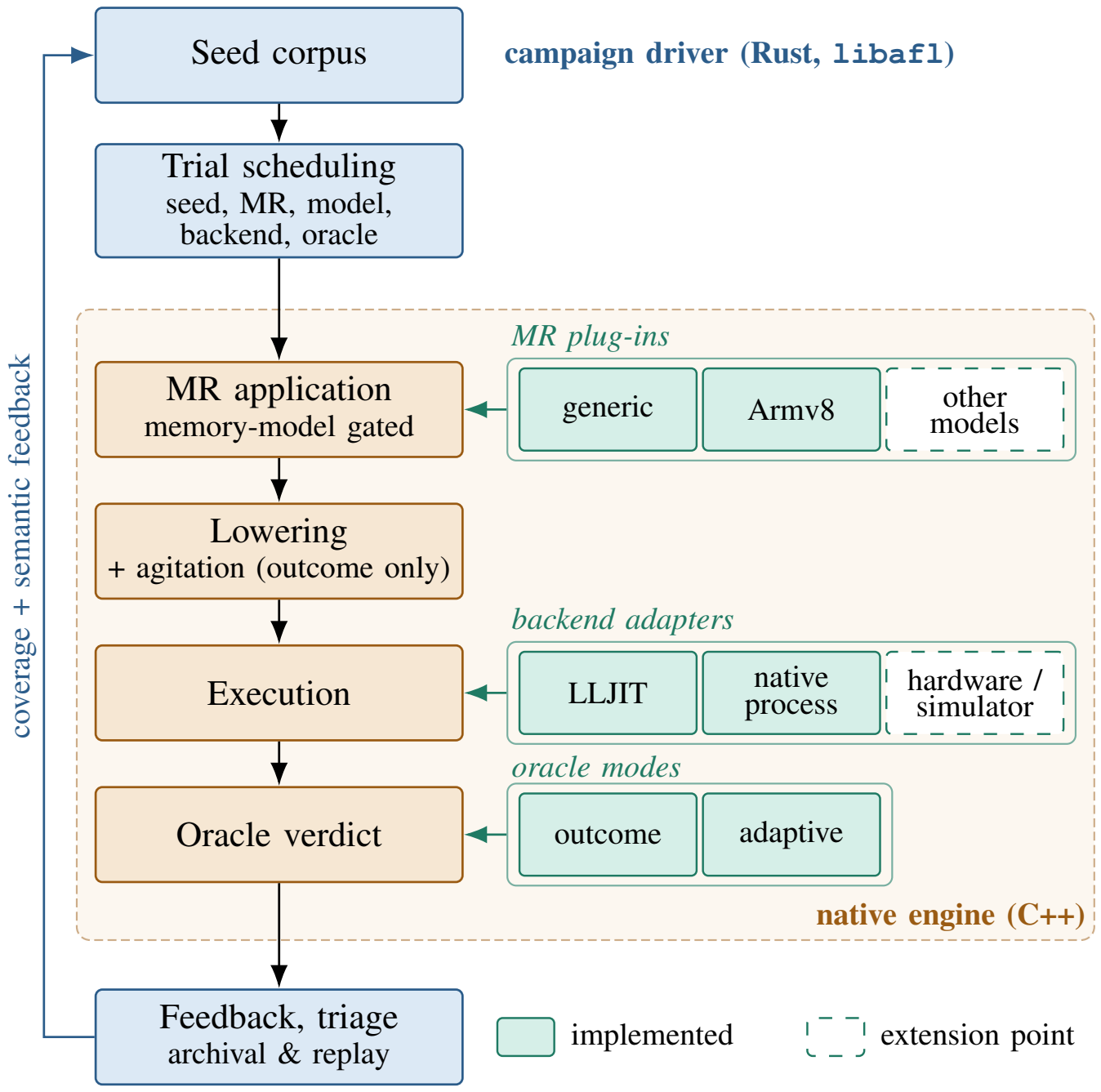

\mlir{}~\cite{mlir} has become the standard shared infrastructure for building multi-level compilers. Its concurrency dialects, such as \openmp{} and \async{}, are lowered through many stages towards targets whose memory models differ from one another~\cite{alglave2014herding, pulte2017simplifying}. Optimisations that appear reasonable can break concurrent programs under weak memory models~\cite{vafeiadis2015common, morisset2013compiler}; such bugs surface as silent miscompilations rather than crashes. Nor is the infrastructure itself immune: whilst developing \tool{}, we identified a bug in the \openmp{} dialect verifier, which failed to reject invalid nested critical operations, allowing programs that hang indefinitely to compile. The bug has since been fixed~\cite{llvm_issue_217354}.

Testing these lowerings is hard. Concurrency amplifies the oracle problem~\cite{barrOracleProblemSoftware2015}: a program's result is a set, or distribution, of outcomes rather than a single value, increasing the amount of reasoning needed. Which outcomes are legal depends on the target, and some arise only rarely -- after many runs or under specific stress conditions~\cite{kirkhamFoundationsEmpiricalMemory2020, alglaveGPUConcurrencyWeak2015}. \mlir{}'s in-tree tests, by contrast, check each program against a singular expected output.
Existing tools do not close this gap. \mlir{} fuzzers~\cite{wang2023mlirsmith, suo2024fuzzing, limpanukorn2025fuzzing, yu2025ratte, liang2025finding} rely on crash, differential or semantic oracles under the assumption of deterministic programs, and their mutations are not memory-model-aware. Where they do produce concurrent programs, these may be racy or have undefined behaviour, so a change in outcome cannot be attributed to the compiler. Concurrency-focused compiler testing~\cite{windsor2021c4, geeson2024compiler, morisset2013compiler} targets \clang{} and specific backends, and often requires a formal model to judge outcomes~\cite{alglave2014herding}. Metamorphic and EMI-based compiler testing~\cite{leCompilerValidationEquivalence2014, donaldson2016metamorphic, donaldsonAutomatedTestingGraphics2017} assumes a deterministic output to compare against. To the best of our knowledge, no tool tests concurrency semantics at the level of \mlir{} without either assuming determinism or requiring a formal model.

We present \tool{}, a metamorphic testing~\cite{chen2018metamorphic, segura2016survey} tool for concurrent \mlir{} programs. Each metamorphic relation (MR) is gated by the target memory model, so that both the original and transformed programs remain well defined. \tool{} judges the outcomes of both programs with a statistical oracle, using either set relations over agitated executions or a comparison of outcome distributions~\cite{guderlei2007statistical}. A coverage-guided campaign driver, built on \libafl{}~\cite{fioraldi2022libafl}, turns this into a continuous fuzzing campaign. \tool{} is aimed at \mlir{} dialect and lowering developers, and at testing researchers who wish to add memory models, MRs, or execution backends; for example, a simulator backend would allow the same campaigns to test system simulators~\cite{even2025search+, dakhama2023searchgem5, dakhamaEnhancingSearchbasedTesting2025}.

In summary, our contributions are:
\begin{enumerate}
    \item \tool{}, an extensible tool for metamorphic testing of concurrent \mlir{}, with four extension points: MRs, memory models, oracle modes, and execution backends.
    \item Campaign infrastructure providing coverage feedback, failure replay, and reproducible runs on \arm hardware.
    \item A case study demonstrating \tool{} at scale on \arm{} hardware, alongside an uncovered bug in \mlir{}'s \openmp{} dialect
\end{enumerate}
\section{\tool{}: Architecture and Design}

\tool{} is a layered metamorphic fuzzer (\Cref{fig:mracle-architecture}).
A \rust campaign driver manages the corpus, schedules trials and records results, while a native \cpp{} engine handles \mlir{}-specific logic, parsing, MR selection, memory-model gating, lowering, execution, and oracle judgement. The two communicate through a narrow C ABI, so the driver is independent of \mlir{} and \llvm{} internals. A trial is fully described by the tuple $\langle \text{seed}, \text{MR}, \text{memory model}, \text{backend}, \text{oracle mode} \rangle$; each element is chosen independently, so the same program pair can be judged under different backends or oracles without changing its MR.

\paragraph{Metamorphic relations}
MRs are implemented as rewrites within a dedicated \mlir{} pass. Each MR in the registry declares the memory model under which it is valid, its rewrite, and the relation it expects between the outcomes of the original and transformed programs: \emph{equality}, \emph{subset}, or \emph{superset}. For each trial, the pass filters the registry by the selected memory model, picks an applicable MR at random, and falls back to another if the rewrite cannot be applied. MRs are either generic (i.e., valid under any model) or model specific. Generic MRs include those intended to increase program diversity, such as \texttt{insert-random-arith}, which insert a random-length chain of \mlir{} \texttt{arith} operations into the program, and concurrency-related MRs that hold under all memory models, such as \texttt{insert-fence}, a subset relation that inserts a fence anywhere valid in a program. In contrast, model-specific MRs are valid only for the memory model they target, such as \texttt{commute-relaxed-read-write}, where relaxed atomic accesses are commuted where the model~\cite{pulte2017simplifying} permits it, which may be unsound under another memory model ~\cite{vafeiadis2015common}. 

\paragraph{Agitation and execution}
A single run of a concurrent program is weak evidence of a fault, or lack thereof, especially since many outcomes are rare~\cite{kirkhamFoundationsEmpiricalMemory2020}. Under the outcome oracle, \tool{} therefore \emph{agitates} both programs before execution. Each program is cloned (10 clones by default), and each clone is perturbed independently: random \texttt{scf.while} jitter at the \mlir{} level, basic block shuffling after lowering to \llvm{} IR, varied optimisation levels, \openmp{} settings and thread counts inspired by traditional litmus testing approaches~\cite{alglave2011litmus}. These perturbations are drawn from the same distribution for both programs, so the relation the MR predicts is preserved. Each clone is then executed repeatedly (2,000 runs by default), and its outcomes are pooled. A cheap single-threaded run of the source program acts as an early exit for deterministic failures. Execution sits behind a backend interface; \tool{} currently provides an in-process LLJIT backend, and a native-process backend. A hardware or system simulator adapter fits the same interface which would enable the same campaigns to test simulators such as gem5~\cite{dakhama2023searchgem5, dakhamaEnhancingSearchbasedTesting2025, even2025search+}.

\paragraph{Oracles}
\tool{} offers two oracle modes. The \emph{outcome} oracle checks the expected relation over outcome \emph{sets}, using a Poisson bound to separate outcomes too rare to support a verdict from genuine violations; inconclusive trials are rerun with freshly agitated clones until they pass, fail, or exhaust a rerun budget. The \emph{adaptive} oracle compares pooled distributions by squared Hellinger distance against a noise envelope obtained by permuting clones between the two sides, so its threshold adapts to the scheduling noise observed. \emph{adaptive} is an instance of statistical metamorphic testing~\cite{guderlei2007statistical}.

\paragraph{Campaign driver}
The driver is built on \libafl~\cite{fioraldi2022libafl}. Each trial runs in a separate worker process that calls the engine, so a crash or hang in \mlir{} is recorded as a finding rather than ending the campaign. Feedback combines edge coverage from instrumented \mlir{} builds with semantic feedback: the MR applied, its expected relation, IR features and the set of observed outcomes. Raw outcome counts are deliberately excluded from novelty, since sampling noise alone would otherwise grow the corpus. Every failure is archived with its transformed program, outcomes, verdict and build and environment metadata, and a replay tool reruns any archived trial outside the campaign. 

\paragraph{Implementation}
\tool{} comprises \textbf{8,594} lines of \cpp, and \textbf{1,583} lines of \rust. A single Cargo build fetches and builds a pinned \llvm{}/\mlir{} (\textbf{373c54d}), the engine and its sanitiser variants. 
\section{Usage}\label{sec:usage}

\paragraph{Installation}
\tool{} runs on Linux and builds with \texttt{cargo build --locked --release} (Rust~1.93.1, Clang~21 with \texttt{libomp}, CMake and Ninja). The first build fetches the pinned \llvm{}/\mlir{} and builds it with coverage instrumentation, and as such, is expensive; later builds reuse it. A \texttt{Dockerfile} packages the same toolchain for users who prefer a container, and prebuilt Docker images are archived on Zenodo~\cite{zenodo}.

\paragraph{Running a campaign}
A campaign is launched with \texttt{mracle-fuzz}, choosing a seed corpus, the memory model whose MRs are enabled (\texttt{--model}; generic MRs are always included), the execution backend (\texttt{--backend}) and the oracle mode (\texttt{--comparison}):
\begin{lstlisting}[language=bash]
target/release/mracle-fuzz \
  --input corpus/seeds --output runs/demo \
  --model armv8 --backend jit \
  --comparison adaptive --duration-secs 86400
\end{lstlisting}
The output directory records the campaign configuration, build and environment metadata, an event log, and a summary, alongside the queue (with provenance for every generated program) and one directory per trial holding its logs and coverage. When a trial fails, its directory also keeps the source and transformed \mlir{}, the lowered \llvm{} IR and the oracle's run information.

\paragraph{Inspecting a finding}
Any trial can be rerun outside the campaign from its saved outputs, which writes a fresh artefact directory:
\begin{lstlisting}[language=bash]
target/release/mracle-replay \
  runs/demo/runs/00000000/request.json \
  --output replay-00000000
\end{lstlisting}

Budgets such as repetitions, run counts, and timeouts can be overridden, e.g. to confirm a rare outcome with more executions.

\paragraph{Extending \tool{}}
Built-in MRs are functions wrapped in a \texttt{FunctionTransform} (listing~\ref{lst:mr}). The function returns \texttt{true} only if the mutation was applied successfully; the entry names the MR, its memory model and its expected relation, which the oracle consumes. MRs that need state can instead subclass \texttt{MetamorphicTransform} and call \texttt{registerTransform()}.
\begin{lstlisting}[language=C++, caption={Adding an MR to a memory model.},
                   label={lst:mr}]
bool tryMyRewrite(func::FuncOp op, RewriterBase &rw,
                  std::mt19937 &rng) {
  // precondition -> safety check -> rewrite
}
const FunctionTransform kMyModelTransforms[] = {
  FunctionTransform("my-rewrite", "my-model",
                    &tryMyRewrite,
                    OutcomeRelation::Equality),
};
\end{lstlisting}
A memory model is simply a named table of this form, exposed through a getter declared in \texttt{Transforms.h} and added to the registry in \texttt{MetamorphicTransform.cpp}; it then becomes selectable via \texttt{--model}. Oracle modes and backends are added in the same way at both layers: a variant in the driver's \texttt{Comparison} or \texttt{Backend} enum, its counterpart in the engine, and a case in the pipeline's dispatch. New seeds are valid \openmp{} \texttt{.mlir} programs placed in \texttt{corpus/seeds}, though this can similarly be extended to other dialects with a \texttt{Backend} extension, such a use would be beneficial for dialect developers wishing to test their dialect's memory model conformance.

\section{Case study}

We ran a small campaign to show that \tool{} operates at scale on real \arm{} hardware, and to give practitioners a sense of the trade-offs between its configurations. 

\paragraph{Setup}
We ran all campaigns on six \arm{} m400 nodes of the Cloudlab testbed~\cite{duplyakin2019design}, each with eight 64-bit ARMv8 (Atlas/A57) cores, and 64GB system memory, with up to four concurrent jobs per node. We crossed the two oracle modes with both backends, giving four configurations, and ran each for 12 hours, three times, with the \arm{} and generic MRs enabled, and a corpus of \textbf{46} seeds, which were manually translated to \openmp{} from the C4 dataset~\cite{windsor2021c4}.

\begin{table}[t]
\centering
\caption{Results per configuration over three 12-hour runs. Trials/h, edges and corpus size are means; findings are totals.}
\label{tab:case-study}
\setlength{\tabcolsep}{4pt}
\begin{tabular}{llrrrr}
\toprule
Oracle & Backend & Trials/h & Edges (k) & Corpus & Findings \\
\midrule
outcome   & JIT    & 2,557  & 93.6 & 1,158  & 10,784 \\
outcome   & native &   846  & 82.2 &   714  &  9,450 \\
adaptive  & JIT    &   471  & 83.3 &   455  &  5,699 \\
adaptive  & native &   162  & 72.0 &   267  &    667 \\
\bottomrule
\end{tabular}
\end{table}

\paragraph{Results}
\Cref{tab:case-study} summarises the 12 runs. Across them, \tool{} attempted 145,296 trials, of which 118,696 completed as normal executions, and 26,600 were flagged as findings. The outcome oracle with the JIT backend was the clear leader, at about 2,560 trials per hour, 5--16$\times$ the rate of the \emph{adaptive} oracle; it also reached the most coverage (93.6k edges) and by far the largest corpus (1,158 programs, against 267–455 for the adaptive oracle). The native backend was around three times slower than JIT under every oracle and reached about 11k fewer edges, although this comparison is not strictly like for like: the JIT backend runs more of the compiler in process, so some of its additional coverage can be attributed to the code required to manage the runtime, rather than the intended exploration.

\paragraph{Findings}
Flag rates differ between the two oracles, and the pattern reverses across backends. Roughly 34\% of adaptive/JIT trials were flagged, against 12\% for outcome/JIT; on the native backend, the adaptive oracle flagged roughly 11\%, against 31\% for outcome. Each finding is a candidate for further triage rather than a confirmed bug: its saved artefacts and replayable output allow a developer to inspect the transformed program and its outcomes directly. Triage of these findings is ongoing; the \openmp{} verifier bug~\cite{llvm_issue_217354} was found during earlier ad-hoc running of the tool during its development.


\paragraph{Limitations}
Three 12-hour repetitions on one platform cannot support statistical claims~\cite{klees2018evaluating}, and coverage is not directly comparable across backends, so we present these results as an indication of use cases rather than an evaluation of effectiveness.
\section{Applications}
Because MRs, memory models, oracles and backends vary independently, \tool{} can be adapted to several groups of users.

\paragraph{MLIR maintainers}
Short outcome/JIT campaigns could run alongside changes to the \openmp{} dialect and its lowerings, with archived trials replayed as self-contained reproducers for bug reports.

\paragraph{Dialect developers}
Developers of new concurrency dialects can supply seeds in their dialect and an execution backend, then use the generic MRs to check that their constructs lower to the behaviour their memory model intends. The JIT backend, which runs lowering and code generation in process, gives fast feedback at this stage.

\paragraph{Target developers}
Teams deploying to targets with weaker or less well-documented memory models can add that model as a table of MRs (\Cref{sec:usage}) and use the native backend, which links against the real \openmp{} runtime, to test the toolchain as deployed. A simulator backend would extend the same campaigns to architectural simulators.

These uses compose: a high-throughput outcome/JIT campaign can generate candidate findings whose inputs seed a slower adaptive campaign, separating likely faults from scheduling noise. We have not evaluated this workflow here.
\section{Conclusion}

We presented \tool{}, a memory-model-aware metamorphic fuzzer for concurrent \mlir{}. By gating MRs on the target memory model and judging outcomes with statistical oracles, it tests concurrency lowering without assuming determinism or requiring complex formal models.
Its MRs, memory models, oracles and backends are independent extension points; future work will add further memory models and a simulator backend, and extend \tool{} beyond the \openmp{} dialect.

\noindent\textbf{Availability:} \tool{}, its seed corpus, campaign scripts and results are available at \url{https://github.com/dakaidan/MLIR-MRacle-Tool}, and archived on Zenodo~\cite{zenodo}.

\bibliographystyle{IEEEtran}
\bibliography{references}

\end{document}